\documentstyle [epsfig,wrapfig,11pt]{article}
\begin{document}
\hfill  ORNL-CCIP-93-11 / RAL-93-065
\vspace{1.0cm}
{
\begin{center}
{\Large\bf Charmonium Physics at a Tau-Charm Factory}
\footnote[1]{Summary talk for the Charmonium Working Group at
the Third International Workshop on a Tau-Charm Factory, Marbella, Spain
(1-6 June 1993).}\\
\vspace{1cm}
T.Barnes\\
Physics Division and Center for Computationally Intensive Physics\\
Oak Ridge National Laboratory, Oak Ridge, TN 37831-6373, USA\\
and\\
Department of Physics and Astronomy, University of Tennessee\\
Knoxville, TN 37996, USA\\

\date{}
\end{center}
}

\begin{abstract}
This talk summarizes the status of the charmonium system, with particular
emphasis on outstanding problems in QCD spectroscopy
which a tau-charm factory can
address.

\end{abstract}

\section{Introduction}
\thispagestyle{empty}
Since its discovery in 1974 the charmonium system has served
hadron physics as an important arena for the investigation of
many aspects of QCD and hadron spectroscopy.
In this summary we briefly review some of these and discuss
several of the important outstanding
issues in hadron spectroscopy and their relation to the
spectrum and couplings of resonances in the charmonium
system. The topics we discuss are charmonium spectroscopy, electromagnetic
couplings ($\gamma$, $\gamma\gamma$ and $e^+e^-$), strong decays and
unusual states (charm
molecules and charmonium hybrids), and in each case we note areas in which
experiments at a tau-charm factory could make valuable contributions.

\section{Charmonium Spectroscopy}

The spectrum of experimental charmonium states \cite{PDG,E7601P1} is shown in
Fig.1,
together with the energy levels predicted by the relativized $c\bar c$
potential model of Godfrey and Isgur \cite{GI}, complete to $\approx 4.2$ GeV.
The experimental states have
$J^{PC}=0^{-+}, 1^{--}, 2^{++}, 1^{++}, 0^{++}$ and $1^{+-}$. Many
$1^{--}$ levels are known since this channel is immediately accessible through
$e^+e^-$ annihilation. One can see that all the experimental resonances
have expected theoretical levels nearby,
with the largest discrepancy being 50
MeV between the observed $\psi(3770)$ and the theoretical
$^3D_1(3820)$.
The overall scheme of levels clearly supports
the presence of a long-range confining
interaction with an asymptotic behavior which is approximately linear.
This allows radial excitations with a slowly decreasing level spacing;
note the masses of the $^3S_1$ candidates, $J/\psi(3097),
\psi(3685), \psi(4040)$ and $\psi(4415)$.
The details of the multiplet splittings support the presence of
short range
one-gluon-exchange interactions, in the L=0 spin-spin interaction
and the splittings of the L=1 multiplet. The L=1 splittings also show
evidence for contributions from an additional, negative, spin-orbit term,
which is expected if the confining
interaction acts as a Lorentz scalar.
Finally, the absence of a significant
long-range spin-spin force, as seen in the
near degeneracy of the S=1 $\chi_j$ multiplet c.o.g. and the
S=0 $h_c(3520)$, is consistent with scalar confinement and argues
against any important vector term.
This L=1 multiplet structure remains the
clearest experimental evidence in support of scalar confinement.

\newpage

\begin{wrapfigure}{r}{4.0in}
\epsfig{file=tcf1.epsm,width=4.0in}
\caption{Charmonium spectrum; experimental resonances [1,2] (lines) and
theoretical $c\bar c$ levels [3] (circles); filled=$1^{--}$, other
$J^{PC}$=open.}
\label{fig 1}
\end{wrapfigure}

Of course
many theoretical levels are predicted which
have not been conclusively identified to date,
such as the non-$1^{--}$ members of the
L=2 multiplet at $\approx 3.8$ GeV, the
radially excited L=1 multiplet
near 3.9 GeV, and other excited-L
levels. The L=2 states $^1D_2$ and $^3D_2$ are
especially attractive experimentally
since they cannot decay strongly to $D\bar D$
and hence should be rather narrow.
The non-$1^{--}$ states are directly accessible in $P\bar P$ annihilation;
indeed, one of the narrow L=2 states may have been observed recently
by the E705 collaboration at Fermilab \cite{E705}.
An L=0 $\eta_c(3590)$ was previously
reported in $\psi(3685)$
radiative decay \cite{XB}, but as this state is not
seen by E760 \cite{Seth} it may not exist at this mass and should certainly
be searched for in $\psi'\to\gamma\eta_c'$ with better statistics.
Although these are interesting experimental targets,
the generally good agreement between experimental and theoretical masses
leads one to ask whether
anything profound
can be learned from future studies of the charmonium system.

In our discussion of transitions and decays we shall see that, despite
the apparent
good agreement
in the energy levels, there are
actually many problems in the
couplings of the states above $D\bar D$ threshold,
and much remains to be understood. It is useful to
divide the charmonium system into a well-understood region near and below
the $D\bar D$ threshold at 3.73 GeV,
and a {\it terra animalium mirabilium}
above $D\bar D$
\cite{AB} where very surprising results have been reported
and new types of states are anticipated by theorists.
In addition to
$|c\bar c\rangle$ basis
states, theorists also expect $|c\bar c g\rangle$ and perhaps
$|c\bar c q_1\bar q_2\rangle$ to be evident in the spectrum,
and physical resonances
will of course be linear superpositions of these states. Hopefully in the
charmonium system this mixing will not be large, so these states can be easily
distinguished. In this review we will use the term ``charmonium" to refer
to all these experimental resonances,
and theoretical assignments such as ``$c\bar c\, $" or
``$c\bar c$-hybrid" are understood to be approximate descriptions of
somewhat more complicated linear superpositions in Hilbert space.

A tau-charm facility should allow us to explore this new territory
above $D\bar D$ threshold and
perhaps answer some of the most interesting outstanding questions in QCD
spectroscopy, including the
possible existence of charm molecules and hybrid mesons.

\section{One-photon transitions}

The experimentally observed single-photon transitions and their partial widths
are shown in Fig.2.
These transitions provide
the only straightforward pathway to many of these levels in $e^+e^-$
annihilation since only $1^{--}$ states are made initially.
These can then
decay radiatively through E1 transitions into $2^{++}$, $1^{++}$ and
$0^{++}$ states.
Similarly we can produce the $^1S_0$,
$0^{-+}$ levels through M1 radiative transitions starting from $1^{--}$
states.
All the
well-established
non-$1^{--}$ charmonium resonances now known except the $h_c(3520)$
were discovered
through these radiative transitions.

Single-photon radiative transitions also provide sensitive tests of proposed
assignments for charmonium resonances, since the rates are proportional
to the squares of wavefunction overlap integrals which can
depend strongly on the
details of the wavefunctions.

There is generally rather good agreement
between theory and experiment in the
$\psi(3685)\to\gamma\chi_j$ and $\chi_j\to\gamma J/\psi$
E1 radiative transitions;
nonrelativistically the amplitudes for these transitions are proportional to
the overlap integrals
$\int d^{\, 3}x\, \Psi^*_i\, r \, \Psi_f$. Most of these E1 rates are
known to an accuracy of about $15\% $. It would probably be unrealistic
to expect greater accuracy from the quark model in the absolute scale, but the
relative rates are subject to less
theoretical uncertainty so it would be useful to
improve these measurements. In the case of the $\chi_j$ decays much of the
statistical error comes from the uncertainty in the total widths, which would
be straightforward to determine more accurately and are also
of theoretical interest.

A more fundamental question in QCD,
the possibility of an anomalous magnetic moment in the
charmed quark's electromagnetic coupling, can also be addressed by
measuring
the photon angular distribution in these transitions.
A study of
$\chi_2\to\gamma J/\psi$ by the E760 collaboration \cite{E760anom}
has found a result consistent with zero anomalous moment, although the errors
are rather large. It would be very interesting to reduce the
uncertainty in this measurement at a tau-charm factory.

\begin{wrapfigure}{r}{4.0in}
\epsfig{file=tcf2.epsm,width=4.0in}
\caption{Observed single-photon transitions and their radiative partial
widths.}
\label{fig 2}
\end{wrapfigure}

For the M1 decays the nonrelativistic transition amplitudes
are proportional to the transition magnetic moment, which is
$e_c/m_c$ times the overlap integral
$\int d^{\, 3}x\, \Psi_i^*(1^{--})\Psi_f(0^{-+})$.
This integral is unity
for states with the same degree of radial excitation and zero otherwise,
if we neglect hyperfine corrections
to the wavefunctions and recoil effects.
Since these rates are suppressed by
$1/(m_c<r>)^2$ they are considerably
weaker than the E1 transitions.
Due to the simple overlap integral the M1 rate for the low-recoil decay
$J/\psi(3097)\to\gamma\eta_c$ should be particularly reliably calculable,
to the extent that we know $1/m_c^2$.
The theoretical expectation for this rate,
$\approx 2.$ KeV (assuming a charm quark mass of
$m_c \approx 1.6$ GeV \cite{GI})
nonetheless does not compare very well with the
experimental 1.1$\pm 0.3$ KeV.
Of course this is only a $3\sigma$ disagreement, but since there is little
systematic uncertainty in the theoretical prediction,
an improved measurement with better
statistics is an important experimental goal.
A more accurate measurement of $J/\psi\to\gamma\eta_c$ would also improve
the results for $\gamma\gamma$ couplings, since
the uncertainty in the background process $J/\psi\to\gamma
\eta_c$ was the dominant systematic error in the recent L3 measurement
of $\eta_c\to\gamma\gamma$ \cite{L3}.
Similarly the radially
excited $\eta_c'$ should be searched for
in $\psi(3685)\to\gamma\eta_c'$, since this transition must
have a radiative partial width near 1 KeV, but E760 has not been able to
confirm the presence of this state near 3592 MeV, as previously reported
in this transition by
the Crystal Ball collaboration \cite{XB}.

\section{Two-photon couplings}

The two-photon couplings of charmonium resonances are measured at $e^+e^-$
machines through the virtual process $e^+e^-\to e^+e^-\gamma\gamma$, with
the two photons subsequently coupling
to a charmonium state, which decays to
the detected
final state. The measurement of the cross sections for these
processes
allows one to infer the strength of the $\gamma\gamma$ coupling
of each resonance, and in the limit of small-$Q_{\gamma}^2$ this
determines the on-shell $\gamma\gamma$ coupling and hence the
$\Gamma_{\gamma\gamma}$ partial width \cite{Bauer}.
Unfortunately the production of charmonium resonances
by this $\gamma\gamma$
process is rather weak, due both to the $e^4$ amplitude and the rapid fall
of effective $\gamma\gamma$ intensity with increasing
$M_{\gamma\gamma}$.
At present the detection of charmonium states in $\gamma\gamma$
is near the limit of
experimental sensitivity. Only the states $\eta_c(2988)$, $\chi_0(3415)$ and
perhaps the $\chi_2(3555)$
have been observed in $\gamma\gamma$ collisions at $e^+e^-$ machines to date,
and these experiments typically report $\Gamma_{\gamma\gamma}$ values of
a few KeV, with errors comparable
to the reported signal.
Fermilab experiment E760 has reported two charmonium $\gamma\gamma$ widths
with rather higher accuracy, using direct hadronic production
of charmonium in $P\bar P$ annihilation
followed by decay to $\gamma\gamma$; the small cross sections
are compensated by high intensity and efficient background
rejection. This approach has led to considerably improved
sensitivity, and in the best case ($\chi_2$) the
statistical error in
$\Gamma_{\gamma\gamma}$ is about an order of
magnitude smaller than at $e^+e^-$ machines.
The experimental partial widths are shown in Fig.3;
note the discrepancy
between the scale of the $\gamma\gamma$
widths reported by $e^+e^-$ facilities
(albeit with rather large errors)
and the E760 results.

\begin{wrapfigure}{r}{3.75in}
\epsfig{file=tcf3.epsm,width=3.75in}
\caption{Observed two-photon transitions and partial
widths.}
\label{fig 3}
\end{wrapfigure}

These partial widths are
interesting as tests of the many quark model predictions for the
$\gamma\gamma$
couplings of $q\bar q$ states; well-known examples are the
nonrelativistic
ratio $\Gamma_{\gamma\gamma}(^3P_0) / \Gamma_{\gamma\gamma}(^3P_2)= 15/4$
and $\lambda=2$ dominance of the $^3P_2$ $q\bar q$-$\gamma\gamma$
coupling.
Recent calculations of relativistic effects
\cite{TBgg} find important corrections in charmonium,
so these couplings can serve as sensitive tests
of relativistic effects if they are measured with
sufficient accuracy. For example, the
$\Gamma_{\gamma\gamma}$ ratio $^3P_0/^3P_2$
for the $\chi_j$ states
is predicted to be
reduced from 15/4 to $\approx 2.8$
by relativistic corrections. This change in ratio should not
be accompanied by any significant $\lambda=0$ production
of the $\chi_2(3555)$, as this is expected to be only
about $0.5\%$ of the $\chi_2(3555)$ $\gamma\gamma$ partial width.
At present
the statistically accurate E760 results appear to support the predictions
of the relativized calculations of $c\bar c$-$\gamma\gamma$ couplings.
An improved measurement of these two-photon partial widths
and angular distributions,
to an accuracy of $\pm 0.030$ KeV or better in
$\Gamma_{\gamma\gamma}^\lambda(\chi_j)$, would allow a sensitive test
of these relativistic
amplitudes and their helicity structure.

In addition to relativistic effects, calculations of
$O(\alpha_s)$ QCD radiative corrections to these $\gamma\gamma$ widths and
other $c\bar c$ transitions
have been reported in the literature.
Unfortunately
these radiative corrections
are renormalization-prescription dependent, so their
numerical importance depends on an unphysical parameter.
Various methods for dealing with this prescription dependence have been
proposed \cite{PD}, which involve calculating
higher order corrections and then choosing the renormalization prescription
for fast convergence in $\alpha_s$ or to minimize sensitivity to the
choice. Typically a large coefficient of $\alpha_s$ is an indication
of an inappropriate choice of prescription.
A more serious problem is that confinement
may modify these gluonic corrections in a
nonperturbative manner, for example through the infrared behavior of the
gluon propagator; this could invalidate
conclusions regarding radiative corrections
which depend strongly
on the
assumption of perturbative gluons at small-$Q_g^2$.

The accuracy of these corrections at a given order in $\alpha_s$
in a given renormalization scheme can easily be tested experimentally.
As an example,
the $O(\alpha_s)$ radiative corrections
to the $\gamma\gamma$ width ratio $^3P_0/^3P_2$ in the scheme advocated
by Kwong {\it et al.} \cite{Kwong} give
$\Gamma_{\gamma\gamma}(\chi_0) /
\Gamma_{\gamma\gamma}(\chi_2) = R_0 \cdot (1 + 0.2\alpha_s/\pi ) /
(1 - 16 \alpha_s / 3\pi)$, which for $\alpha_s=0.3$ changes the nonrelativistic
ratio of $R_0=15/4$  to about 7.8, corresponding to
$\Gamma_{\gamma\gamma}(\chi_0) = 2.5$ KeV using the E760
$\Gamma_{\gamma\gamma}(\chi_2)$ width.
In contrast, with only relativistic corrections
we expect
$\Gamma_{\gamma\gamma}(\chi_0) = 0.9$ KeV; obviously it is
straightforward to test
these two theoretical results through accurate
measurements of $\Gamma_{\gamma\gamma}(\chi_0)$ and
$\Gamma_{\gamma\gamma}(\chi_2)$. The total width ratio
$\Gamma(\chi_0)/
\Gamma(\chi_2)$ can be used similarly.
These tests of QCD radiative corrections at the charmonium
mass scale can obviously
have wide implications regarding the
range of applicability of perturbative QCD.

The L$>$1 $c\bar c$ states are expected to have very
weak $\gamma\gamma$ couplings, so they may remain experimentally
inaccessible in this process.
For example,
in recent calculations
the $^1D_2$ expected at $\approx 3.8$ GeV is predicted to
have $\Gamma_{\gamma\gamma}\approx 30$ eV \cite{TBgg}. Earlier
nonrelativistic calculations gave a somewhat larger estimate of
$140-200$
eV \cite{Novikov}, although this reference overestimates
$\Gamma_{\gamma\gamma}(\chi_2)$ relative to the accurate E760 result.
Production of
radially excited $c\bar c$ states from $\gamma\gamma$ is not expected
to be suppressed significantly, so $\gamma\gamma$ may serve as
a filter for L=0 $0^{-+}$ and L=1 $0^{++}$ and $2^{++}$ $c\bar c$ states above
$D\bar D$ threshold,
given adequate statistics. Radial excitations are especially interesting
because none have yet been identified in $\gamma\gamma$ production
of light $q\bar q$ systems, despite
predictions of unsuppressed $\gamma\gamma$ couplings.
It is not clear if this is a problem for theory, because light radials
with these quantum numbers are poorly understood,
and their branching fractions to the final states reconstructed
to date
are unknown and may be small.

Previous $e^+e^-$ experiments have exclusively used the radiative process
$e^+e^-\to e^+e^-R$ to determine $\gamma\gamma$ couplings.
Another possibility \cite{Seth}
which might be exploited at a tau-charm facility is the annihilation process
$e^+e^-\to \psi(3685) \to \gamma\chi_j$, $\chi_j\to\gamma\gamma$. These
$\psi'\to\gamma\gamma\gamma$ branching
fractions
are $\sim 10^{-5}$, so for a plausible sample
of $10^8$ $\psi'$ events at a tau-charm factory
we would have about $10^3$
$\chi_j\to\gamma\gamma$ decays, and could then determine
$\Gamma_{\gamma\gamma}(\chi_j)$ to an accuracy of a few
$\%$. This is sufficient to
allow sensitive tests of the relativistic
and radiative corrections cited above.

\section{$e^+e^-$ couplings}

The final electromagnetic process we consider is single-photon
production
of $1^{--}$ states, which is measured directly in $e^+e^-$
annihilation. These partial widths
are shown in Fig.4, and have errors of typically $\pm 0.2$ KeV.
Nonrelativistically we would expect production of only $^3S_1$ states,
since the nonrelativistic production amplitude is proportional to $\Psi(0)$.
There are nonlocal relativistic corrections to this result, however,
so some
production of $^3D_1$ states is also expected,
with an amplitude proportional to
$\Psi''(0)$ nonrelativistically.
In $c\bar c$ potential models this
$^3D_1$ amplitude is
much weaker than the $^3S_1$ coupling.
For example, in the Godfrey-Isgur model one expects
$\Gamma_{e^+e^-}(\psi(3770))/\Gamma_{e^+e^-}(J/\psi(3097)) \approx 0.010$,
assuming that these states are dominantly $^3D_1$ and $^3S_1$ respectively.

As we can see in Fig.4, the $e^+e^-$ coupling of the $\psi(3770)$ is indeed
much weaker than that of the L=0
states $J/\psi(3097)$ and $\psi(3685)$.
However the observed magnitude of the coupling disagrees
with theoretical expectations; the ratio
$\Gamma_{e^+e^-}(\psi(3770))/\Gamma_{e^+e^-}(J/\psi(3097))$ is about
a factor of five
larger than Godfrey and Isgur predict. (The $\psi(3770)$ errors are actually
rather large and should be improved.)
This may be due to a
$^3S_1$ component in the $\psi(3770)$, which might be tested by
a determination of its E1 transition rates to the L=1 $\chi_j$ states.
Such an admixture is driven by the tensor term in the one-gluon-exchange
Hamiltonian, but this effect is already incorporated in the Godfrey-Isgur model
and the mixing is not large enough to explain the observed $e^+e^-$ partial
width.
In earlier, closely related work Eichten {\it et al.} \cite{Eichten} found that
the $e^+e^-$ width of the $\psi(3770)$ could be explained by
$^3S_1$-$^3D_1$ mixing through virtual $D\bar D$ intermediate states.
The size of
such virtual meson-pair effects is an important and currently rather
obscure issue,
and studies at a tau-charm factory may
clarify this issue,
for example through a more accurate determination of the composition of
the nominally $^3D_1$ states $\psi(3770)$ and $\psi(4160)$.

\begin{wrapfigure}{r}{3.5in}
\epsfig{file=tcf4.epsm,width=3.5in}
\caption{Experimental $e^+e^-$ partial widths of $1^{--}$
charmonium resonances.}
\label{fig 4}
\end{wrapfigure}

For the
higher-lying $1^{--}$ states
these $e^+e^-$ couplings are even more problematical;
the $\psi(4160)$ is usually considered a radially excited $^3D_1'$ $c\bar c$
state
due to its mass, but it has an $e^+e^-$ coupling comparable to
those of the putative $^3S_1^{(n)}$ states
$\psi(4040)$ and $\psi(4415)$. Either there is very
important configuration mixing or the $\psi(4160)$ has been misidentified.
A detailed
scan of R from $D\bar D$ threshold to the highest accessible energy
at a tau-charm factory should
clarify the spectrum of
$1^{--}$ $c\bar c$
resonances.
This will also
be an important contribution to the identification of non-$c\bar c$
$1^{--}$ states such as
charmed-meson molecules and charmonium hybrids, since these may
exist in this mass range
and will only be apparent
once the conventional $1^{--}$ $c\bar c$
states have been identified.
The presence of these additional states may
account for some of the unusual properties
reported for the higher-mass
$\psi$ resonances.

\section{Strong decays: $c\bar c$ versus charm molecules}

Below the $D\bar D$ threshold of 3.73 GeV most
hadronic decays of charmonium
states involve annihilation into
light hadrons. This is usually approximated by
$c\bar c$ annihilation into free gluons,
$ggg$ for $^3S_1$ states and $gg$ for $^1S_0$, $^3P_2$ and $^3P_0$.
(See especially \cite{Novikov} and \cite{Eichten} for these results.)
The qualitative ordering of strong widths is accounted for by this
approximation, although higher-order strong corrections to these rates
are problematical and can appear quite large; see our discussion of
radiative corrections to $\gamma\gamma$ widths in this regard. There are
interesting problems in the exclusive hadronic final states,
for example the $\rho\pi$ branching fraction
from the $J/\psi(3097)$ is at least two orders of magnitude larger
than from the $\psi(3685)$. Actually, much of the interest in these
annihilation decays
is
not in the annihilation process itself, but is instead
due to the possibility of detecting gluonic states
\cite{revs} such
as glueballs or light hybrids in the final state.

Above $D\bar D$
most charmonium resonances
can couple to open-charm
decay channels, and they
have hadronic widths in the 10s of MeVs.
Remarkably little is known about
the branching
fractions of the four charmonium resonances reported above 3.73 GeV.
(See Figs.1 and 5.)
The
$\psi(3770)$ is reported \cite{PDG}
to decay dominantly to $D\bar D$, but ``hairpin diagrams" which
would allow cascade
processes such as $\psi''\to J/\psi\pi\pi$ are known to be weak
(these contribute a partial width of $\approx 100$ KeV to
$\psi(3685)$ decays)
so this is hardly surprising.
Of the two highest levels, nothing is claimed
for the strong modes of the $\psi(4160)$, and the $\psi(4415)$ is reported to
decay dominantly to hadronic final states, again not a surprise.

The single resonance
above $D\bar D$ threshold with known branching fractions
is the $\psi(4040)$, and the experimental
results are puzzling. The branching fractions to
$D^*\bar D^* : D^* \bar D$+$h.c.: D\bar D$ are in the ratios
$32\pm 12 : 1 : 0.05\pm 0.03$.
Thus, the dominant mode is $D^*\bar D^*$, despite the proximity of the
$\psi(4040)$ to the $D^*\bar D^*$
threshold of 4.02 GeV.
The $D^*\bar D^*$ mode is
reported to be stronger than $D\bar D$ by about
three orders of magnitude, despite the much smaller $D^*\bar D^*$ phase space,
which is a surprising result indeed. These branching fractions
are inferred from the assumed resonant part of the cross section
$e^+e^-\to M_1M_2$, in which the ratios are less extreme due to spin
multiplicity factors. The reported $\psi(4040)$ branching fractions
may be biased
by misidentified nonresonant production of charmed meson pairs,
and should be remeasured with improved accuracy and careful determination of
the nonresonant background.
This nonresonant production is
expected to contribute to the cross section for $e^+e^- \to M_1M_2$
in the ratio $7:4:1$ \cite{FEC}.

In view of these remarkable branching fractions,
Voloshin and Okun \cite{molec}
suggested that the $\psi(4040)$ might be a $D^*\bar D^*$
molecular state,
similar to the $K\bar K$-molecule description of the $f_0(975)$ and $a_0(980)$
proposed
subsequently by Weinstein and Isgur \cite{WI}.
This is an obvious suggestion given
the mass and branching fractions reported for the $\psi(4040)$,
although
it does not explain why this state has an $e^+e^-$ coupling about equal to
expectations for a $^3S_1''$ $c\bar c$ state, which is
anticipated near this mass. In recent work
Ericson and Karl \cite{EK} concluded that one pion exchange forces are
sufficiently strong to bind the $D^*\bar D^*$ system,
although they find a
$^1S_0$, $J^P=0^+$ ground state rather than $1^-$.
Of course both molecular and $c\bar c$ states may exist in this mass region.

Since the $^3S_1''$
$c\bar c$ assignment for the $\psi(4040)$ is a second radial
excitation, there are two zeroes in the strong decay
amplitude as a function of $|\vec P_f|$, and
the $D\bar D$
and $D^*\bar D$+$h.c.$
modes may have been ``accidentally" suppressed
by their values of $|\vec P_f|$. This possibility
was investigated in the $^3P_0$ model by LeYaouanc
{\it et al.} \cite{ccbar},
who concluded that the observed decay modes did indeed arise
naturally from nodal suppression of the decay amplitudes.
This $c\bar c$
model can be tested in future through measurements of the
branching fractions of the other candidate radially excited
$c\bar c$ states such
as the $\psi(4415)$.
Of course the $^3P_0$ and other $q\bar q$ pair-production decay
models are very
phenomenological, and
a better understanding
of couplings to open channels may be required to explain the
branching fractions of these higher-mass states.
At present, models
typically find important $c\bar c$ mass shifts
due to couplings to open channels \cite{open},
which is surprising in view of the success of naive $c\bar c$ potential
models in explaining charmonium spectroscopy.
Accurate experimental
studies of strong decays
should
lead to considerable refinement of the models and hopefully to an understanding
of why these effects appear small below $D\bar D$ threshold.

These branching fractions will be of great interest
for studies of charm physics as well,
because future
experiments on charmed mesons such as the $D_s$ will
presumably use the
higher-mass
$\psi$ resonances as charmed-meson factories,
and the branching fractions
will determine the optimum $\psi$-resonance
source of each charmed meson.

\section{Charmonium Hybrids}

The search for gluonic excitations in the hadron spectrum may be the most
interesting topic in QCD spectroscopy \cite{revs}.
Hadrons which contain
quarks and excited glue
in their dominant basis states are referred to as ``hybrids", and in the
meson sector these can have exotic-$J^{PC}$ quantum numbers which are forbidden
to $q\bar q$ states.
The charmonium system provides a natural laboratory for the study of hybrids
because the ordinary $c\bar c$ spectrum is rather straightforward and, with
better statistics above 3.8 GeV, can probably be clarified considerably.
It should then be
possible to identify any additional states such as hybrids or charm molecules
against the background of $c\bar c$ resonances.
Since charm molecules are expected to lie just below two-meson thresholds
and have S-wave quantum numbers,
there should be little confusion between these
two types of non-$c\bar c$ states.

The masses of light hybrid mesons have been estimated using
the MIT bag model \cite{hybag},
QCD sum rules \cite{hysr},
the
flux tube model \cite{hyft} and
heavy-quark lattice gauge theory \cite{hylgt}.
Although there is considerable variation in detail, all these approaches
predict
that the
lightest hybrid mesons have masses of $\approx 1.5 - 2.$ GeV, and in
heavy-quark systems (from the flux-tube model and lattice gauge theory)
near the lightest $Q\bar Q$ mass plus $\approx 1$ GeV.
The exotic quantum
numbers $J^{PC}=1^{-+}$ are often suggested for experimental searches
in light-quark systems,
because all techniques find a light hybrid with these quantum numbers, and
the flux tube model predicts that the I=1 $1^{-+}$
should be relatively narrow. This model finds
an especially
rich lowest-lying hybrid multiplet, with $J^{PC} = 1^{\pm\pm}, 2^{\pm\mp},
1^{\pm\mp}$ and $0^{\pm\mp}$ all approximately degenerate. The mass estimated
for this lowest $c\bar c$-hybrid multiplet has varied
between 4.19 GeV and 4.473 GeV in flux tube references \cite{hyft}.
Perantonis and Michael
\cite{hylgt} find 4.04 GeV for the lightest hybrids
in heavy-quark
lattice gauge theory
in the quenched approximation, and estimate 4.19 GeV
without this approximation.
Finally, Narison \cite{hysr}
quotes a QCD sum rule result of 4.1 GeV for the $1^{-+}$ exotic
$c\bar c$-hybrid, consistent with the lattice gauge
theory and lower flux-tube results.
Theoretical estimates
of the mass of the lightest $c\bar c$-hybrid multiplet
are thus typically about $4.2 \pm 0.2$ GeV.

Theoretical models predict rather characteristic
two-body decay modes for hybrids.
In both constituent gluon \cite{hycg} and flux tube \cite{hyft} models
the lightest hybrids are found to decay preferentially
to pairs of one L$_{q\bar q}$=0
and one L$_{q\bar q}$=1 meson,
for example $\pi f_1$ and $\pi b_1$. These unusual modes
have received little
experimental attention and may explain why hybrids were not discovered
previously. For $c\bar c$-hybrids this implies that the hybrid mass
relative to the S+P threshold of $\approx 4.3$ GeV is an important issue;
if hybrids lie below this threshold their preferred decay modes will be
closed, and they will be correspondingly narrow and detection may be
straightforward.
Hybrids below this threshold are preferred
by most models, and in this case the dominant modes will probably be
$D^{(*)} \bar D^{(*)}$, with weaker contributions from cascade decays to
$c\bar c$ plus light hadrons. Prospects for searching these modes for hybrids
are discussed below and by Close \cite{Close2}.

There are several possible strategies for producing
hybrid charmonium states at a tau-charm factory.

First, one may search for additional $1^{--}$ states directly
through a high-statistics scan of R. Although
hybrids are expected
to appear weakly since they must couple through their
$c\bar c$ components, they might nonetheless be evident as small, relatively
narrow peaks which
can be excluded
as conventional $c\bar c$ resonances by their masses and quantum numbers.
(Of course the narrow width is speculative,
based on the closed S+P channel for
$c\bar c$-hybrid masses below $\approx 4.3$ GeV
and the preferred-mode argument.)
This experiment is straightforward since a detailed scan in R would be
an early priority at a tau-charm factory in any case.

\begin{wrapfigure}{r}{3.5in}
\epsfig{file=tcf5.epsm,width=3.5in}
\caption{Production of a $c\bar c$-hybrid by cascade decay from
the $c\bar c$ continuum.}
\label{fig 5}
\end{wrapfigure}

Instead of producing a $1^{--}$ hybrid directly one can search for hybrids
in the decay products of initial $1^{--}$ $c\bar c$ resonances or the
$c\bar c$ continuum.
Ideally a search for charmonium hybrids would concentrate on the
$J^{PC}$-exotic states, to preclude confusion with $c\bar c$.
Indeed, since several excited-L $c\bar c$ multiplets are expected in this
mass region (see Fig.1) it may be important to identify these as well.
Decay to a hybrid state $H_c$ can occur
through a strong cascade
decay such as $(c\bar c) \to \eta H_c$ or
$(c\bar c) \to \pi\pi H_c$.
Since the branching fraction of each cascade may be $\sim 10^{-3}$,
the large numbers of events expected at a tau-charm factory will be
required to make detection of hybrids practical using this mechanism.
In these cascade decays
a selection rule may help identify hybrids; the $H_c$ and the
light hadronic system ($\eta$ or $\pi\pi$) will be produced
in a relative
P-wave. This origin of this selection rule can be visualized in the
related process of cascade decay of a flux-tube hybrid; this
occurs by
``pinching off" a loop of glue from the excited flux tube,
which initially had an $\exp(i\phi)$ wavefunction about the
$c\bar c$ axis. The final
$c\bar c$ state has a ground-state flux tube, so the $\exp(i\phi)$ dependence
must be transferred to
the $c\bar c$-(light hadronic) relative orbital wavefunction, which
therefore must have L$\geq$1. We assume that the decays are dominated by
the minimum value L=1.
An interesting suggestion
\cite{Bugg} is that this search can involve a continuum cascade;
the initial $c\bar c$ system can be produced in the high mass
continuum, at perhaps $\sim 5$
GeV, and will then cascade into a $c\bar c$-hybrid plus a light hadron
or hadrons.

Detection of the final $H_c$ hybrid can follow two approaches; either
it can be reconstructed from conventional pair-production final states
such as $D\bar D$ or $D^*\bar D$, or it can be found in a second cascade
decay.
These options are suggested in Fig.5.
In the double cascade, which is attractive due to the simplicity of the
final states involved,
the $H_c$ can cascade into $J/\psi$ plus a hadron or
hadrons. One would
select events such as $\eta\eta J/\psi$ from the 5 GeV $c\bar c$ continuum,
and search for new states in the $\eta J/\psi$ invariant mass distribution
and for exotic quantum numbers in the angular distributions.
The cascade selection rules discussed above
suggest a search for $H_c\to \eta J/\psi$ (P-wave) and
$H_c\to (\pi\pi)_S J/\psi$ (P-wave, with the $\pi\pi$ system in S-wave).
This $\eta J/\psi$ system can have $J^{PC}=2^{--},1^{--}$ and
$0^{--}$ ($0^{--}$ is predicted to be an excited-state
exotic hybrid in the flux tube model),
and $(\pi\pi)_S J/\psi$ in P-wave
can have $2^{+-}, 1^{+-}$ and $0^{+-}$;
$2^{+-}$  and $0^{+-}$ are ground-state exotic hybrids in the flux tube
model.
We would also expect conventional $c\bar c$ states to be made in
cascade, albeit dominantly in S-wave combinations
with the light hadrons; one could search for the
$h_c'(\approx 3960)$ for example in the $\eta J/\psi$ S-wave in the same
experiment.

\section{Summary and Conclusions}

In this review we have discussed the experimental status of charmonium
and listed interesting topics for experimental investigation
at a tau-charm factory. The general areas for charmonium-related experiments
and what we might learn from them are as follows:

\vskip 0.5cm
\noindent
1) {\it Spectroscopy above $D\bar D$ threshold.} Are gluonic excitations
evident
in the charmonium spectrum? Models predict additional hybrid-charmonium states
starting at about 4.2 GeV, some with exotic quantum numbers.
These states may stand out clearly in a scan of
R or in cascade decays (from high-mass $c\bar c$ states or the
$c\bar c$ continuum).
Is there evidence for charm-meson molecules, or can all observed levels be
attributed to $c\bar c$ states or perhaps $c\bar c$-hybrids?

\vskip 0.5cm
\noindent
2) {\it Strong decays
of resonances above $D\bar D$ threshold.} Little is known about these,
and the reported branching fractions for the $\psi(4040)$ are remarkable.
Measurements of the branching fractions of higher-lying resonances will
allow tests of decay models, which will clarify the status of
possible charm molecules. These measurements will also suggest optimum
sources for production
of the various charm mesons (such as the $D_s$) which may be of interest
in weak interaction physics.

\vskip 0.5cm
\noindent
3) {\it Electromagnetic couplings of charmonia.} The $\gamma$,
$\gamma\gamma$ and $e^+e^-$
couplings of charmonium resonances can discriminate between a wide range of
theoretical predictions in the literature, and can test the validity of
relativistic correction formalisms and
the applicability of perturbative QCD
radiative corrections at the charmonium mass scale.
There are many open questions in these electromagnetic couplings
which can be resolved through
accurate measurements at a tau-charm factory.
Especially interesting
are the M1 decays $J/\psi\to\gamma\eta_c$ and
$\psi'\to\gamma\eta_c'$ and E1 transitions such as
$\chi_2\to\gamma J/\psi$ (which can
test for an anomalous $c$-quark magnetic moment), $\gamma\gamma$
couplings of the $\{ \chi_j\} $ and
$\eta_c'$, and the $e^+e^-$ couplings of candidate
$^3D_1$ $c\bar c$ states.

\vskip 0.5cm

Clearly, a very rich program of charmonium physics is possible
at a tau-charm factory, and many problems in spectroscopy which have been
unresolved since the 1970s can be addressed at this facility. We strongly
advocate its approval.

\section{Acknowledgements}

This contribution is a summary of discussions and material presented by the
Charmonium Working Group in parallel session at the Third Workshop on the
Tau-Charm Factory. The participants in the working group were T.Barnes,
N.Bartolomeo, D.Bugg, F.E.Close, M.Doser, A.Falvard, R.Landua,
J.Lee-Franzini,
A.Palano and K.K.Seth. I
would like to thank them for their participation and contributions
to this summary. I would also like to thank P.Geiger,
S.Godfrey, N.Isgur, G.Karl, M.Shifman
and P.M.Stevenson
for additional discussions of material presented here.
It is a pleasure to acknowledge
J.Kirkby and associates
for their invitation to co-organize and chair the
charmonium working group for the
tau-charm meeting. The assistance of my colleague F.E.Close is
also gratefully acknowledged. This research was sponsored in
part by the United States Department of Energy under contract
DE-AC05-840R21400, managed by
Martin Marietta Energy Systems, Inc, and by the United Kingdom Science Research
Council through a Visiting Scientist grant at Rutherford Appleton Laboratory.

\end{document}